\documentclass[aps,prx,twocolumn,showpacs,superscriptaddress,groupedaddress]{revtex4-1}  % for review and submission
\usepackage{amssymb,amsmath,graphicx}
\usepackage{graphicx,wrapfig,lipsum}
\usepackage[usenames,dvipsnames]{xcolor}
\usepackage{graphicx}  % needed for figures
\usepackage{dcolumn}   % needed for some tables
\usepackage{bm}        % for math
\usepackage{amssymb}   % for math
\usepackage{braket}
\usepackage{mathtools}
\usepackage{hyperref}
\usepackage{xcolor}
\hypersetup{
    colorlinks,
    linkcolor={blue},
    citecolor={blue},
    urlcolor={blue}
}
\newcommand{\vect}[1]{\mathbf{#1}}

\usepackage{appendix}
\usepackage{ulem}
\newcommand{\bsf}[1]{\textsf{\textbf{#1}}}

\begin{document}
\title{Waves, Algebraic Growth and Clumping in Sedimenting Disk Arrays}
\author{Rahul Chajwa $^1$, Narayanan Menon $^{2}$, Sriram Ramaswamy $^{3}$, Rama Govindarajan $^{1}$}
\affiliation{1. International Centre for Theoretical Sciences, Tata Institute of Fundamental Research, Bengaluru 560 089 \\ 
2. Department of Physics, University of Massachusetts, Amherst MA 01003 USA \\ 
3. Department of Physics, Indian Institute of Science, Bengaluru 560 012 }

%\date{\today}
%\maketitle
\begin{abstract}
An array of spheres descending slowly through a viscous fluid always clumps [J.M. Crowley, J. Fluid Mech. {\bf 45}, 151 (1971)]. We show that anisotropic particle shape qualitatively transforms this iconic instability of collective sedimentation. In experiment and theory on disks, aligned facing their neighbours in a horizontal one-dimensional lattice and settling at Reynolds number $\sim 10^{-4}$ in a quasi-two-dimensional slab geometry, we find that for large enough lattice spacing the coupling of disk orientation and translation rescues the array from the clumping instability. Despite the absence of inertia the resulting dynamics displays the wavelike excitations of a mass-and-spring array, with a conserved ``momentum'' in the form of the collective tilt of the disks and an emergent spring stiffness from the viscous hydrodynamic interaction. However, the non-normal character of the dynamical matrix leads to algebraic growth of perturbations even in the linearly stable regime. Stability analysis demarcates a phase boundary in the plane of wavenumber and lattice spacing, separating the regimes of algebraically growing waves and clumping, in quantitative	agreement with our experiments. Anisotropic shape thus suppresses the classic linear instability of sedimenting sphere arrays, introduces a new conserved variable, and opens a window to the physics of transient growth of linearly stable modes.

\end{abstract}
\maketitle  
\section{ Introduction}
The collective settling of particles in viscous fluids is a  classic and notoriously difficult problem in the physics of strongly interacting driven systems. 
In the Stokesian limit of Reynolds number $Re \to 0$, inertia is negligible, viscous forces dominate, and a settling particle creates a flow field that decays slowly with distance $r$ as $1/r$ \cite{Stokes1851, HB1, kim1, brady}. Furthermore, particles in most natural and industrial settings are not spheres, and the hydrodynamics of settling couples their rotational and translational degrees of freedom \cite{jeffery1, chwang, witten1, haim}.
The separation vector of two sedimenting spheres is constant \cite{jeffery2,smoluchowski}, thanks to Stokesian time-reversal symmetry  \cite{purcell}. That of a pair of spheroids, in sharp contrast, is either time-periodic or asymptotically diverging \cite{kim2,wakiya, shelley}, an effect usefully understood through a surprising and precise analogy to Kepler orbits \cite{chajwa}. How particle anisotropy transfigures many-body sedimentation \cite{segre,SR1,ladd,guazzelli}  is the central theme of this work.

\begin{figure}[t]
\begin{center}
\includegraphics[width=8.5 cm]{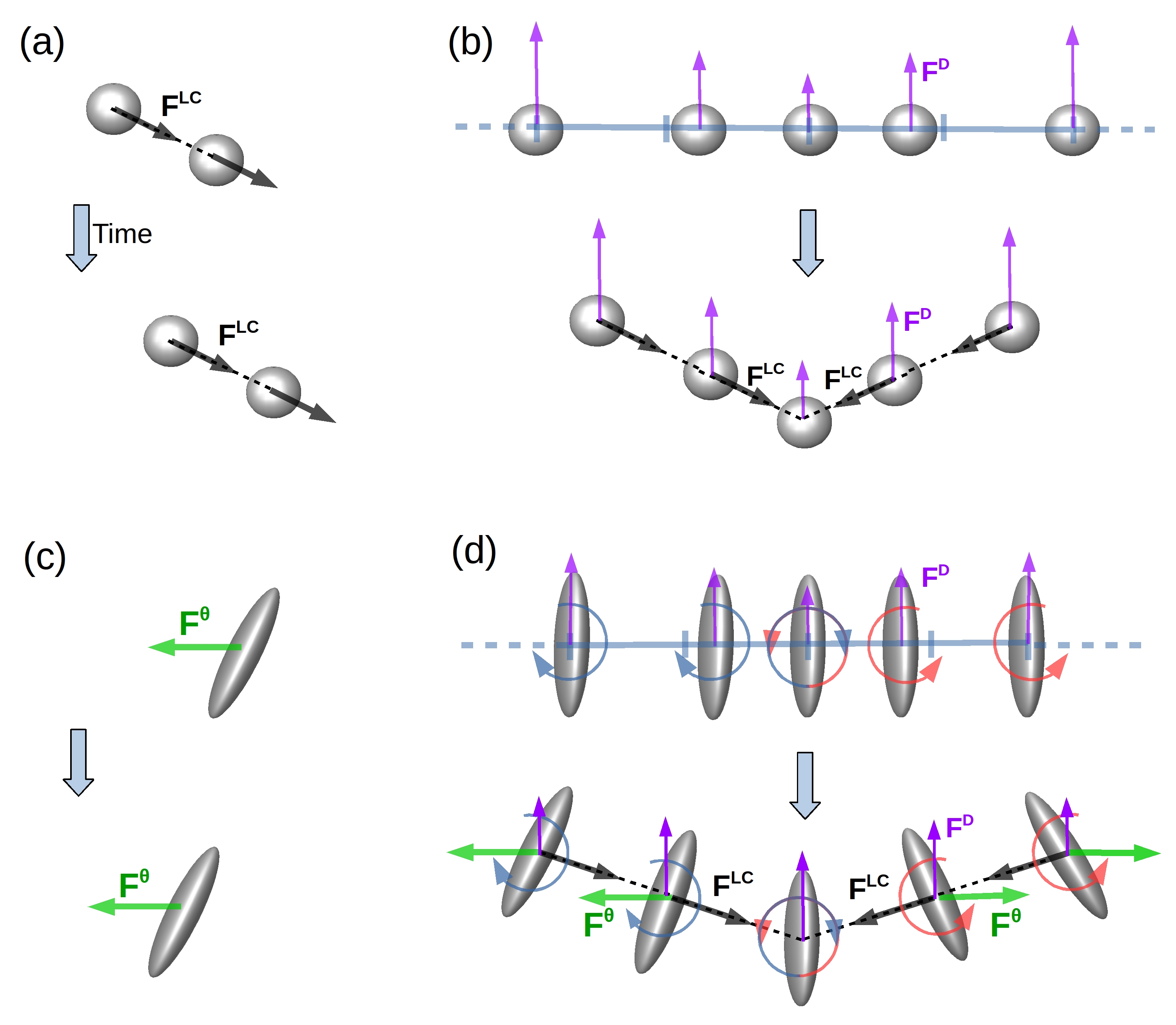}
%\hspace{0cm}\newline
\caption{\label{Fig1}\textbf{Schematic of competing mechanisms}: (a) $F^{LC}$ acts on the pair along line joining their centers leading to horizontal drift. (b) Clumping of array of spheres result from the line of centres force $F^{LC}$, acting along with reduced drag force $F^{D}$. (c) A spheroid drifts laterally as it falls when it is oriented obliquely with respect to gravity (d) Lateral drift $F^{\theta}$ competing with the clumping induced by $F^{LC}$ and $F^{D}$. }
\end{center}
\end{figure}

The statistics of number fluctuations in sedimentation have been studied for collections of apolar \cite{KS,guazzelli2,tornberg} and polar \cite{witten2} anisotropic particles in a steady state with spatially uniform mean concentration. Sedimenting lattices, on the other hand, break translation invariance and thus retain a reference microstructure \cite{zick} about which they display a rich dynamics \cite{crowley,crowley2,sfm,LR, LBR, simha} distinct from that of the uniform suspension. In addition, the presence of a lattice clarifies the connection between particle-level interactions and long-wavelength collective phenomena as seen in Crowley's celebrated clumping instability \cite{crowley, crowley2} of a regular array of sedimenting Stokesian spheres with purely hydrodynamic interactions. The Crowley instability can be understood by the composition of two-body interactions: (i) a trio of particles packed slightly closer than the rest settle faster due to reduced drag $\vect{F}^{D}$; (ii) the resulting local tilt of the array leads to a lateral drift force $\vect{F}^{LC}$, acting along the line joining their centers \cite{HB1} [see Figure \ref{Fig1} (a) - (b)]. $\vect{F}^{D}$ and $\vect{F}^{LC}$ together lead to dense regions breaking away from the array in clumps, on a scale given by the wavelength of the initial perturbation [see Supplementary video 1]. In this paper we ask how non-spherical shape alters this central and inescapable feature of the sedimentation of sphere arrays.

We pursue this question experimentally and theoretically through the simple yet unexplored case of a freely sedimenting linear array of orientable apolar particles. Such a particle in isolation, aligned obliquely and settling under gravity, drifts laterally [see Figure \ref{Fig1} (c)] with a velocity that depends, for a given orientation, on the particle geometry through a mobility function \cite{HB1} whose analytical form is known for spheroids \cite{chwang,kim2}. When a collection of such particles settle in an array, the lateral drift $\vect{F}^{\theta}$ of an individual particle with tilt angle $\theta$ can compete with the line-of-centers force on a pair of particles $\vect{F}^{LC}$,  potentially preventing the clumping instability [see Figure \ref{Fig1} (d)]. We therefore ask: is a sedimenting lattice of oriented objects stable? We answer this question for a system of disks, that is, oblate spheroids with eccentricity $e \to 1$, as they display the most pronounced lateral drift \cite{chwang,kim1,kim2}. We note further that, despite their ubiquity in nature, the sedimentation of disk-like objects is much less studied than that of their rod-like counterparts \cite{guazzelli2,tornberg}, We show later that our findings can be generalised to all uniaxial apolar shapes.

In our experiments, we impose initial positional perturbations on a configuration in which the disks stand face to face, that is, each disk with its face normal aligned with the vectors joining it to its nearest neighbours, in a uniformly spaced horizontal array as depicted in Fig.\ref{Fig1}(d). In the plane of perturbation wavenumber $q$ and lattice spacing $d$, we find two distinct dynamical regimes --  wavelike excitations with algebraic growth, with no counterpart in sphere arrays, and linearly unstable modes -- separated by a stability boundary. Crucially, the observed algebraic growth of perturbations occurs outside the regime of linear instability, in the neutrally stable regime. Using symmetries, we show that the long-wavelength dynamics of an array of uniaxial apolar objects, with the short dimension initially aligned horizontally, contains terms that compete with Crowley's \cite{crowley} clumping instability. 
 
Explicit construction of the dynamical equations of motion for Stokesian sedimenting spheroids, at the level of pair hydrodynamic interactions, determines the values of coefficients in our coarse-grained theory, and accounts for the experimentally observed instability boundary in the $q$-$d$ plane. In the neutrally stable regime we find an emergent elasticity which formally resembles that of a mass-and-spring chain, with the orientations of the disks playing the role of a momentum density field which is conserved when summed over the entire lattice. 
The mode structure of the linearised nearest-neighbour theory, in the limit of thin disks, compares remarkably well with the experimentally measured frequency $\omega$ of the waves, wherein $\omega \to 0$ as wavenumber $q \to 0$. This ``hydrodynamic'' character of the modes is a dual consequence of translation invariance along $x$ as a result of which only the relative $x$ positions of the disks matter, and the apolar character of the disks, as a result of which there is no restoring torque if all disks are rotated through the same angle, and so the sum of all the angles acts like a conserved total momentum.

We observe transient algebraic growth of perturbations in the linearly stable regime in our experiments, and in the numerical solution of the far-field equations.  We term this growth ``nonmodal'' since it occurs even when all modes of the dynamical matrix are neutral or decaying \cite{farrell,schmid,trefethen,bale}. The underlying reason is that our dynamical matrix $A$ is nonnormal, i.e., $AA^\dagger \ne A^\dagger A$ (where the dagger represents the adjoint). Once the perturbation amplitude due to this nonmodal growth is large enough, nonlinearities can be triggered, disrupting the lattice through an unconventional route to instability at late times. Our calculations further predict the form of the initial perturbation that leads to maximum transient growth at each point in the neutrally stable regime in the $q$-$d$ plane. 

In following sections, we first show our experimental findings and rationalise them heuristically. Then we present a symmetry-based coarse-grained description, which is followed by the construction of the dynamical equations and their quantitative comparison with experimental observations.

\section{Experiments}
\begin{figure}[b]
\begin{center}
\includegraphics[width=8.5cm]{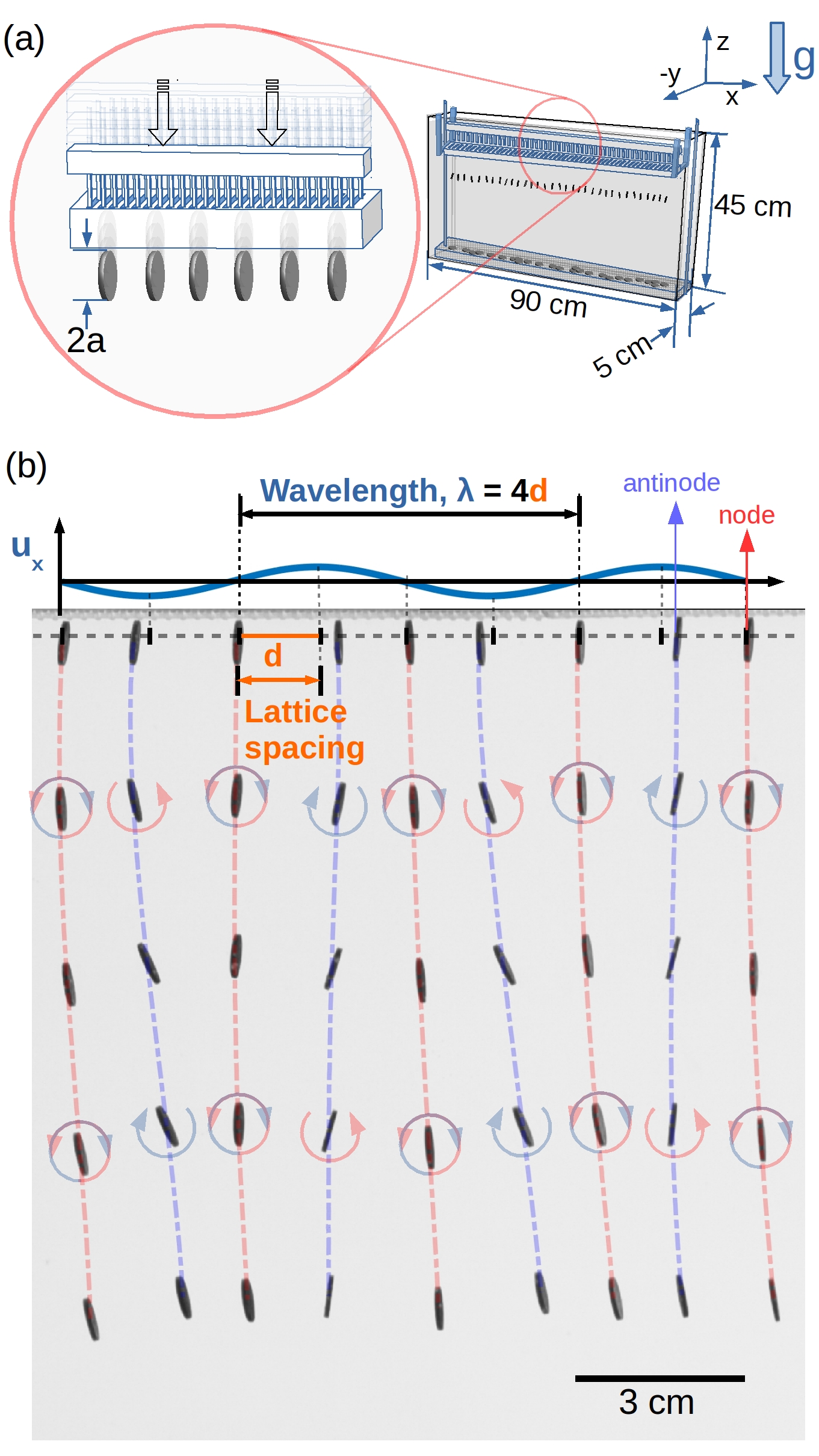}
%\hspace{0cm}\newline
\caption{\label{Fig2} (a) Schematic of the experimental setup. (b) \textbf{Linearly stable mode}: Overlapped time frames shown for time, t = 0, 130, 258, 386, 516 sec. The unperturbed lattice of initial conditions is shown in gray dashed lines. Horizontal positional perturbation, $u_{x}$ is sinusoidal with wavelength $\lambda = 4 d$, lattice spacing of unperturbed state  $d=3.75a$, leads to undulations in orientation and vertical positions. The trajectories of disks at nodal and antinodal points is given by red and blue dashed lines respectively. The rotations of disk orientations in clockwise and anti-clockwise directions is given by blue and red circular arrows respectively. The sense of rotation changes along antinodal trajectories [see Supplementary video 2].}
\end{center}
\end{figure}

Our experiments were conducted with disks of radius $a$=0.4 cm and thickness 1 mm, 3D printed (FormLabs SLA) with stereolithography using resin of density 1.164 g cm$^{-3}$, settling in Silicone oil of density 0.98 cm$^{-3}$ and kinematic viscosity $5000$ cSt leading to a typical Reynolds number $Re \sim 10^{-4}$.  The particles are released in a one-dimensional array from the top of a quasi-two-dimensional glass container of width ($x$-direction) = $225 \, a\,$, height ($z$-direction) =  $\, 112.5 \, a \,$, and depth ($y$-direction) = $  \, 12.5 \, a $. As shown in Figure \ref{Fig2} (a), the disks were initially placed with their surface normals $\vect{K}$ perpendicular to gravity and parallel to the line joining the disk centres. This was achieved by first placing the disks in  slots separated with a centre-to-centre spacing of $0.625a$ within a frame centred along the depth of the container. The clearance of the disks in the slot sets a precision of $0.0625 a$ in the horizontal position,  a deviation of up to $1.8  ^{\circ} $ in orientation from the vertical, and negligibly small differences in initial vertical positions.  Both this frame and the disks are already submerged in the fluid to suppress air bubbles. The disks are then ejected from the slots at the same time with a comb whose teeth fit the slots in the array. The centres of the discs and their surface normals lie in the central $(x,z)$ plane of the experimental geometry for much of their trajectory.

\begin{figure}[b]
\begin{center}
\includegraphics[width=8.5cm]{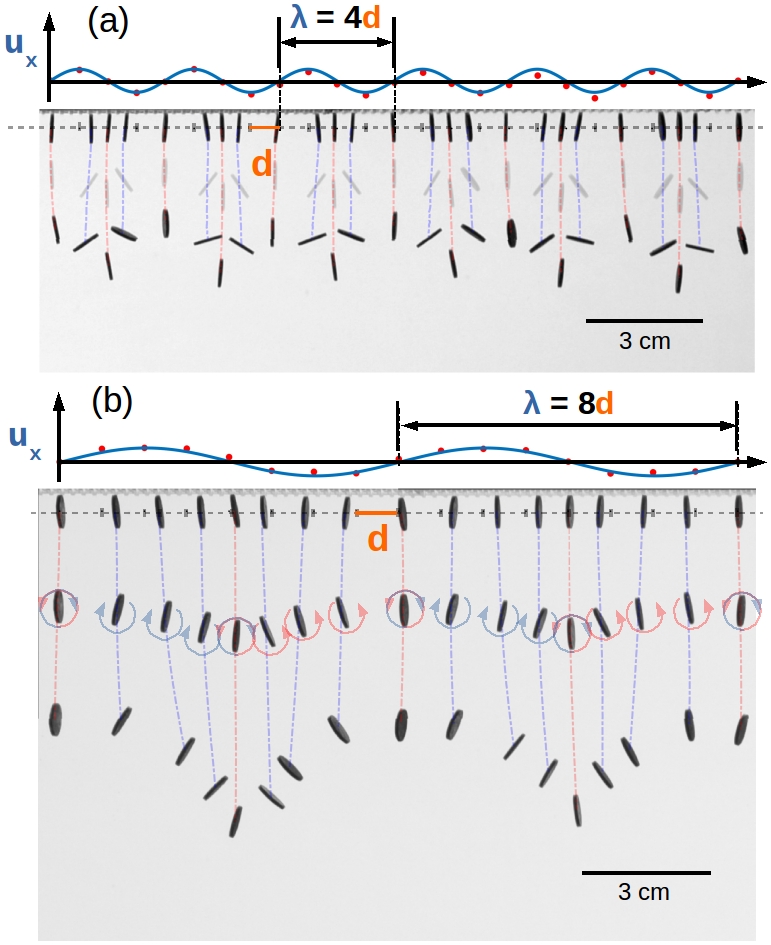}
%\hspace{0cm}\newline
\caption{\label{Fig3}\textbf{Clumping instability}: Overlapped frames (a) t =0, 72, 135 seconds, and (b) t= 0, 120, 255 seconds. In (a) initial horizontal perturbation wavenumber $q = 2\pi/\lambda = \pi/2d$ [same as Fig (\ref{Fig2}b)] and lattice spacing $d=1.875a$, exhibits an unstable case, as line of centers force $F^{LC}$ dominates over orientational drift $F^{\theta}$ . In (b)  $d=2.5a$ and $qd = \pi/4$, exhibits an unstable case where $F^{LC}$ dominates over $F^{\theta}$ , leading to coarsening of the lattice followed by clumping in non-linear regime.  The trajectories of disks at nodal points is given by red dashed lines. Clockwise and counterclockwise rotation of disks, depicted as blue and red circular arrows respectively, do not change colour along the trajectory of the disks, in contrast with the wave shown in Fig (\ref{Fig2}b) [see Supplementary videos 4 \& 5].}
\end{center}
\end{figure}

The reference state is an array of disks with uniform spacing achieved by choosing slots with separation $d$.  On top of this initially uniform lattice, we impose horizontal positional perturbations $u_x(t=0)$ at a wavenumber $q$ by displacing disks appropriately to the right or left slot [see Appendix Figure \ref{fig:AA1}]. The initial perturbations were measured to be sinusoidal with good accuracy, despite the discrete nature of the horizontal displacement.  The clearance of the disks in the slots leads to random errors in initial orientations, which contributes to an error of $0.06 a$ in imposed positional perturbation [see Appendix Figure \ref{fig:AA3}]. 

Images were taken at 1/3 frame per second using a Nikon D750 D-SLR camera. The positions and orientations of the disks were tracked by fitting ellipses to every disk for each image frame. The centroid and angle of the ellipse give the centre  positions and orientation of the disks respectively, with precisions of $0.02 a$ and $0.5 \deg$. The time-dependent amplitude of the positional and orientation perturbation ($u_x, u_z, \theta$)  were measured by fitting a sine wave to the measured particle displacements and orientation relative to the reference lattice in the co-moving frame.

\section{Two regimes of dynamics}
As we vary lattice spacing, $d$, and perturbation wave-vector $q$,  we experimentally observe two distinct regimes of dynamics, as depicted in Figure \ref{Fig2}(b) and \ref{Fig3}.

(i) Waves of orientation coupled with number density fluctuations ---  for the initial condition in Figure \ref{Fig2} (b), we see that the disks at the density nodes hardly rotate, while the orientation and position of disks at the antinodes vary sinusoidally with time  [see Appendix figure 7].  Qualitatively, these wave dynamics may be explained by a composition of drag reduction, horizontal glide and mutual rotation as discussed in Figure \ref{Fig1}(d). Disks in regions of high number density fall faster than those in less dense regions due to reduced drag. The translational degree of freedom couples with rotations such that the disks in the dense region spread out due to orientational glide, stabilising the lattice. This mechanism is characterized by change in sign of the rotation of disks at the antinodal points, which leads to waves [see Supplementary video 2]. This wave is eventually disrupted [see Supplementary video 3], due to an amplification by nonmodal growth mechanism of the experimental imprecision in the initial orientation [see Appendix Figure \ref{fig:AA3}], as shown later in this article.

(ii) Clumping instability decorated with orientations  --- a different type of dynamics is observed for the initial conditions in Figure \ref{Fig3} where the perturbation quickly sharpens at the displacement nodes, or the high density regions. Just as in the Crowley instability of spheres, the dense regions fall faster due to reduced drag, and the vertical perturbation $u_{z}$ increases. The orientation acts to spread out and rarefy the dense regions, but this effect is suppressed by the line of centers force leading to a clumping instability. The rotation of the antinodal points does not change sign, in contrast with the wave-like regime. Figure \ref{Fig3} (a) depicts a marginally unstable case where the initial horizontal perturbation neither grows nor decays substantially whereas in Figure \ref{Fig3} (b) the horizontal perturbations grows to make dense region more dense [see Supplementary video 4 \& 5]. 

Later in the article, we show experimentally the regime of each of these two types of dynamics by varying initial conditions in the $q$-$d$ space. 

Theoretically, we go beyond our qualitative explanation above at two levels. First, we understand our experimental observations using symmetries of the equations of motion for displacements and orientations, in the continuum limit of our system. Second, to determine the phenomenological coefficients in the symmetry-based equations for Stokesian sedimentation, we construct the dynamical equations of the lattice using pairwise addition of forces and torques resulting from the hydrodynamic interactions. We then show that the linearized dispersion relation of our theory compares quantitatively well with our experiments, while long time nonlinear instabilities can be understood by numerical investigations of the far-field equations of motion. 

\section{Sedimenting spheroid lattice: symmetry-based continuum theory}
We construct the ``hydrodynamic'' equations of motion of a drifting lattice of orientable objects, in the limit of no inertia, by writing the most general form of the mobility tensor (defined by velocity = mobility $\times$ force) allowed by the symmetries of the system, to leading order in a gradient expansion, extending theories \cite{LR,LBR} of the statistical dynamics of sedimenting crystals of pointlike objects. We find that the dynamical response of a lattice of orientable particles when perturbed about a suitable reference state contains terms that can compete with the clumping instability of isotropic particles \cite{LR,crowley}. We discuss the structure of the resulting wavelike modes. 

The configurations of a periodic lattice of uniaxial objects are characterized, in a coarse-grained Eulerian description, by the displacement field $\vect{u}$ of the lattice and the orientation field $\vect{K}$ defined by the mean local alignment of the particle axes. For our geometry [see Figure \ref{Fig4} (a)] $\vect{K} = (\cos \theta, 0, \sin \theta)$, with $\theta$ equivalent to $\theta + \pi$ because the particles are fore-aft symmetric. The equations of motion for $\vect{u}$ and $\vect{K}$, in the presence of a gravitational driving force $\vect{F}$, must obey the following symmetries:\\
\begin{itemize}
  \item Stokesian time-reversal symmetry under $t\to -t$ and $\vect{F} \to -\vect{F}$ \cite{HB1}
  \item Translational invariance
  \item Rotational invariance in the subspace perpendicular to gravity
  \item Symmetry under inversion of orientations, $\vect{K} \to - \vect{K}$
\end{itemize} 
The mobility cannot depend directly on $\vect{u}$ due to translational invariance, but dependence on $\nabla \vect{u}$, $\vect{K}$ and $\nabla \vect{K}$ is allowed:
\begin{equation}
\frac{\partial \vect{u}}{\partial t} = \bsf{M} (\nabla \vect{u}, \vect{K}, \nabla \vect{K}) \cdot \vect{F},
\label{eqn:1}
%\dot{u} = M_{ij}^{R} F^{j}
\end{equation}
\begin{equation}
\frac{\partial \vect{K}}{\partial t} = \bsf{P} \cdot \bsf{N}  (\nabla \vect{u}, \vect{K}, \nabla \vect{K}) \cdot \vect{F},
\label{eqn:2}
%\dot{K}_{i} = M_{ij}^{\theta} F^{j} 
\end{equation}
where $\bsf{M}$ and $\bsf{N}$ are the translational and rotational mobilities respectively, and $\bsf{P} \equiv \bsf{I} - \vect{K}\vect{K}$ is the projector transverse to the unit vector $\vect{K}$. The other symmetries further constrain the allowed form of translational and rotational mobilities [see Appendix B], leading, at lowest order in gradients,
for a one-dimensional lattice along $x$, in a comoving frame, to
\begin{equation}
\frac{\partial u_{x}}{\partial t} \, =  \, \lambda_{1} \frac{\partial u_{z}}{\partial x} \, + \, \alpha K_{x} K_{z},
\label{eqn:5}
\end{equation} 
\begin{equation}
\frac{\partial u_{z}}{\partial t} \, =  \, \lambda_{2} \frac{\partial u_{x}}{\partial x} \, + \, \beta K_{z}^{2},
\label{eqn:6}
\end{equation}
\begin{equation}
\frac{\partial K_{z}}{\partial t} = \gamma K_{x} \, \frac{\partial ^{2} u_{x}}{\partial x ^{2}}.
\label{eqn:7}
\end{equation}
Here, $\lambda_{i}$, $\alpha$ and $\gamma$ depend on $\vect{F}$ and the parameters governing the mobilities in \eqref{eqn:1} and \eqref{eqn:2}. Note: equations \eqref{eqn:5} - \eqref{eqn:7} contains only hydrodynamic couplings proportional to the gravitational driving force. We have not included interactions arising from interparticle potentials or entropy. These enter at next order in gradients, and break Stokesian time-reversibility \cite{purcell}. Substituting $\vect{K} = (\cos \theta, 0, \sin \theta)$ in  \eqref{eqn:5}-\eqref{eqn:7} and linearizing about $\theta = 0$, the state where the particle axes are along $x$ (Fig. \ref{Fig2}a) leads, for disturbances with frequency $\omega$ and wavenumber $q$, to the dispersion relations    
\begin{equation}
\omega_{0} = 0,\quad \omega_{\pm} = \pm q_{x} \sqrt{\lambda_{1}\lambda_{2} + \alpha\gamma}
\label{eqn:8}
\end{equation}
with elasticity contributing to \eqref{eqn:7} and \eqref{eqn:8} at order $q^2$. For $\alpha \to 0$ the linearized equations for the translational degrees of freedom $(u_x,u_z)$ are independent of $\vect{K}$ and reduce to those of the LR model \cite{LR}, with wavelike modes or an instability depending on the sign of $\lambda_{1}\lambda_{2}$ \cite{LR,LBR,simha}. $\vect{u}$ affects $\vect{K}$ through the one-way coupling governed by $\gamma$. For $\alpha \neq 0$, translation and rotation are coupled, and the presence of $\alpha\gamma$ in the dispersion relation opens up the possibility of linearly stable wavelike dynamics even for $\lambda_{1}\lambda_{2}<0$. The linearized dynamics about the state where $\vect{K}$ is vertical corresponds to changing the sign of $\alpha$ in \eqref{eqn:8}. For a system of sedimenting particles this means that the array is stable either with horizontal orientations or vertical orientations, but not both. Similar considerations arise in principle for the stability and dynamics \cite{simha} of driven flux lattices in thin slabs of type-II superconductors if the cross-sections of the flux lines are non-circular. 

\section{Sedimenting Spheroid lattice: Pair hydrodynamic interactions}
We now go beyond symmetry considerations, and explicitly construct the equations of motion for a settling lattice based on single-particle motion and addition of pairwise forces and torques at each particle position. We develop the theory for an array of spheroids, of eccentricity $e = \sqrt{1 - b^{2}/a^{2}}$, where $a$ and $b$ are the  semi-major and semi-minor axes respectively. In the limit of $e \to 1$, an oblate spheroid approaches a disk shape, as in our experiments. We consider hydrodynamic interactions to leading order in $a/r$, where $r$ is the separation between two particles. The ingredients of array dynamics are: \\
(i) Lateral drift of a single particle -- An isolated settling spheroid cannot rotate, thanks to Stokesian time-reversal symmetry, but drifts horizontally with velocity
\begin{equation}
U_{x}^{0} = \frac{F \alpha(e)}{12\pi \mu a} \sin 2\theta
\label{eqn:9}
\end{equation}
\cite{HB1,chwang} where $F$ is its buoyant weight, $\mu$ is the dynamic viscosity of the fluid and the mobility $\alpha$ is a function of the eccentricity. Figure \ref{Fig4}(a) shows a schematic of a portion of our array, in which the orientation vector $K^{n}$ of the $n^{th}$ particle is defined as a unit vector along the minor (major) axis for an oblate (prolate) spheroid. The angle $\theta^n$ is measured from the vertical as shown.

(ii) Mutual drag reduction --  Two particles at finite separation fall faster than an isolated one, due to the addition of the flow fields generated by each Stokes monopole \cite{HB1,crowley}. In the far-field approximation, the increased vertical velocity to leading order in $a/r$ is
\begin{equation}
U_{z} = - \frac{F}{6\pi \mu a} \frac{3a}{4 r}\left[1 + \frac{(z_{1} - z_{2})^{2}}{r^{2}}\right].
\label{eqn:10}
\end{equation}
(iii) Horizontal drift --  The flow generated by the neighbouring particle gives rise to a force along the line joining the centers of the two particles \cite{HB1,crowley}, which leads to a horizontal component of velocity 
\begin{equation}
U_{x} = -\frac{F}{6\pi \mu a} \frac{3a}{4 r^{3}}  (x_{1} - x_{2})(z_{1} - z_{2})
\label{eqn:11}
\end{equation}
to leading order in $a/r$.
 
(iv) Mutual rotational coupling -- The presence of a neighbouring particle generates a velocity field of non-zero vorticity, which to leading order in $a/r$ gives a rotation
\begin{equation}
\dot{\theta} = F \frac{x_{1}-x_{2}}{8\pi\mu r^{3}}.
\label{eqn:12}
\end{equation}
We combine ingredients (i) to (iv) to build the dynamics of the array of spheroids.
 
\subsection{ Mode structure for oblate and prolate spheroids}
\begin{figure}[h]
         \begin{center}
         \includegraphics[width=8.5cm]{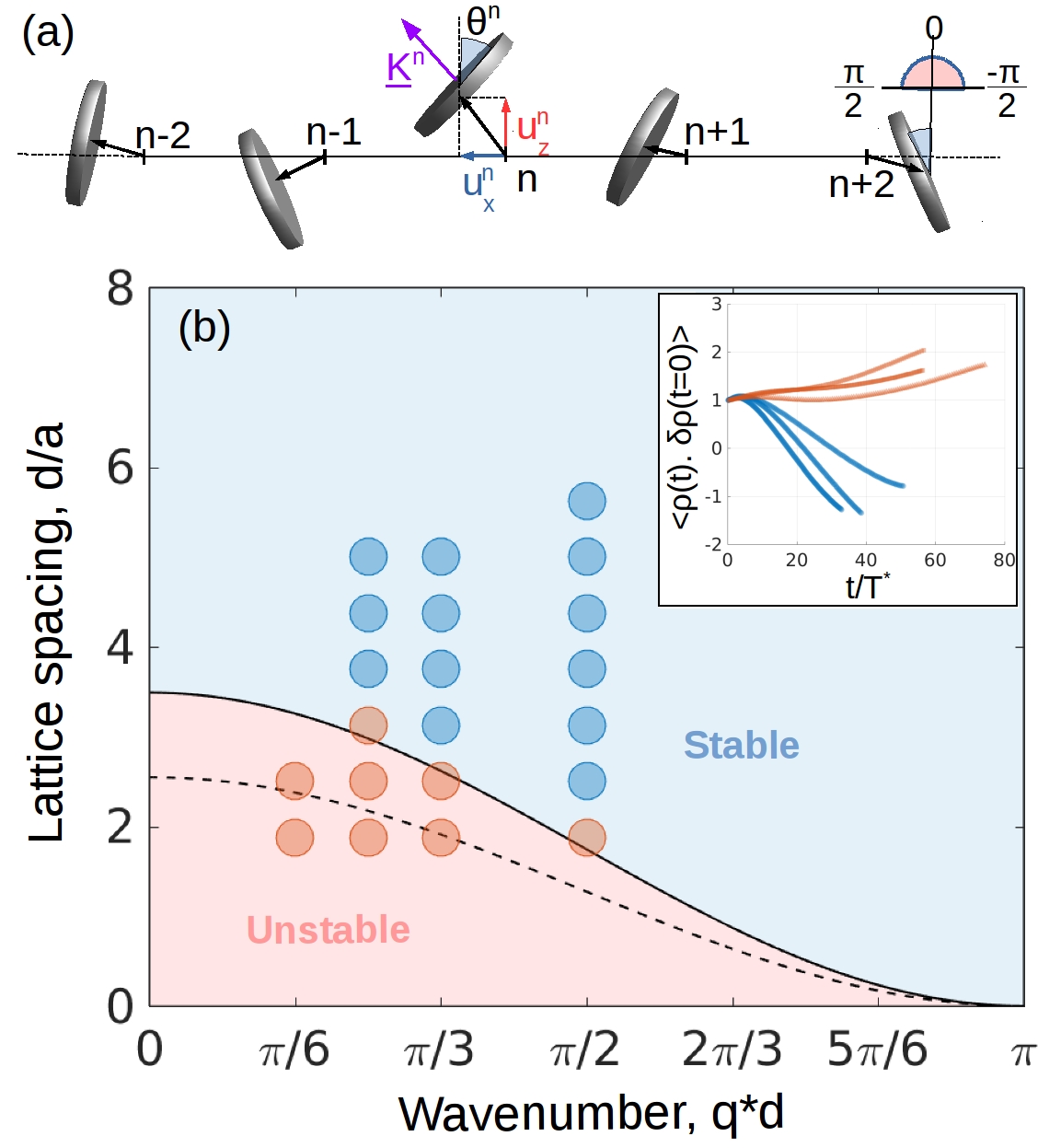}
         %\hspace{0cm}\newline
         \caption{\label{Fig4}\textbf{Phase diagram}: (a) A schematic of the array of disks showing spatial perturbations ($u^{n}_{x}$ ,$u^{n}_{x}$) and orientation perturbation $\theta^{n}$ of the $n^{th}$ disk which interacts hydrodynamically with the neighbours $n-i$, $i = 1, 2, .. $ . For prolate spheroids the orientation vector $\vect{K}^{n}$ is rotated by $\pi/2$ from the one shown. (b) The phase diagram with the stable regime shown in blue and unstable regime in red. The experimental data points (circles) are coloured blue or red by measuring whether the density autocorrelation grows or decays, which is shown in the inset to (b) on the top right, for some representative stable (blue) and unstable (red) points in the $q -d$ plane. The phase boundary predicted by the linear theory with nearest-neighbour interaction is shown for oblate spheroids with eccentricity approximating the experimental thickness, $e =0.9922$ (solid line) and and elliposid of zero thickness $e=1$ (dashed line). }
         \end{center}
         \end{figure}

We consider an infinite  one-dimensional reference lattice along the $x$-axis of  uniformly spaced lattice points with spacing $d$ and falling in the $-z$ direction. As shown in Figure \ref{Fig4}(a), we consider identical spheroids with orientation $\theta^{n}$, and centroids at a small displacement $(u^{n}_{x},u^{n}_{z})$ measured from each lattice point, where the superscript $n$ stands for the $n^{th}$ particle.  In the mean settling frame, pairwise addition of forces and torques on the $n^{th}$ particle due to hydrodynamic interactions with the $(n+l)^{th}$ and $(n-l)^{th}$ particles, for $l = 1,2,3 ... \infty $, gives the equation of motion of the $n^{th}$ particle as
 \begin{equation}
 \frac{d {u_{x}}^{n}}{dt} = -\frac{F}{8 \pi \mu }  \sum_{l=1}^{\infty}  \frac{{u_{z}}^{n+l} - {u_{z}}^{n-l}}{l^{2}d^{2}} \, + \, \frac{F \alpha (e)}{12 \pi \mu a} \sin 2\theta^{n},
 \label{eqn:13}
 \end{equation}
 \begin{equation}
  \frac{d {u_{z}}^{n}}{dt} = \frac{F}{8 \pi \mu} \sum_{l=1}^{\infty}  \frac{{u_{x}}^{n+l} - {u_{x}}^{n-l}}{l^{2}d^{2}} + \frac{F \alpha (e)}{6 \pi \mu a} \sin ^{2}\theta^{n},
  \label{eqn:14}
  \end{equation}
  \begin{equation}
    \frac{d \theta^{n}}{dt} = -\frac{F}{4 \pi \mu} \sum_{l=1}^{\infty} \frac{{u_{x}}^{n+l} + {u_{x}}^{n-l} - 2{u_{x}}^{n} }{l^{3}d^{3}}.
    \label{eqn:15}
   \end{equation}
   
For $u_{z} =0$, equations \eqref{eqn:13} and \eqref{eqn:15} governing the dynamics of the settling array are formally identical to those for the displacement and momentum-density fields respectively of a momentum-conserving lattice of masses and springs. Note that the orientation plays the role of momentum and there is a resultant conservation of the total momentum, $\sum_n\theta^n$ in \eqref{eqn:15}. For fixed $F$ the equations of motion are invariant under $t \to -t$, $u_x^n \to u_x^n$, $\theta^n \to -\theta^n$, $u_z^n \to -u_z^n$.  A term in \eqref{eqn:15} of the form $\theta^{n+1} + \theta^{n-1} - 2 \theta^n$, which within our analogy amounts to a momentum-conserving viscous damping, can arise if inter-disk entropic or energetic aligning interactions, which break Stokesian time-reversal invariance, are taken into account. We do not pursue this issue further here except to note that within a linear stability analysis such a term would turn a neutral regime into a stable one.

In a quasi two-dimensional geometry the dynamics can be approximated by a nearest-neighbour treatment,  where the $n^{th}$ particle interacts hydrodynamically only with the $(n+1)^{th}$ and $(n-1)^{th}$ particle. We non-dimensionalise equations \eqref{eqn:13} - \eqref{eqn:15} using the lattice separation $d$ and time scale $T^{*}= \mu d^{2}/F$, and perturb the angle $\theta =0 + \delta \theta$. Linearising the equations and fourier transforming gives the equation  $\dot{\vect{X}}_{q} = \vect{A}(q) \vect{X}_{q}$, where $\vect{X}_{q} = (u^{x}_{q}, u^{z}_{q}, \delta \theta_{q}) $ is the spatial fourier transform of the perturbations with wavenumber $q$ along $x$, with a nonnormal dynamical matrix: \\
   \begin{gather}
     A(q)
     = \left(
      \begin{array}{ccc}
       0 &  -i \sin(q)/4\pi & \alpha(e)d/{6\pi a} \\
        i\sin(q)/4\pi & 0 & 0 \\
       -\sin^{2}(q/2)/{\pi} & 0 & 0
       \end{array} 
       \right).
       \label{eqn:16}
    \end{gather}
 We return to the interesting consequences of the nonnormality of $A(q)$ later.  For now, we substitute the translational mobility function \cite{kim1,kim2,chwang} for oblate spheroids: $\alpha(e) = \left[\left(9-6 e^2\right)  \tan ^{-1}\left({e}/{\sqrt{1-e^2}}\right)-9 e
    \sqrt{1-e^2}\right]/{8 e^3}$, and for prolate spheroids: $\alpha(e)= { \left\{\left(9 - 3 e^2\right) \ln \left[{(e+1)}/{(1-e)}\right]- 18 e\right\}}/{16 e^3}$, which gives the mode structure with two branches around $\omega =0$ for each:\\
  \begin{widetext}
  
  \text{Oblate Spheroids:}
\\  
  \begin{equation}
  \quad i \omega _{\pm}(q) = \pm \frac{1}{8\sqrt{3} \pi e^{3/2}  } \sqrt{\sin ^2\left(\frac{q}{2}\right) \left( \frac{12 d}{a}{\left(2 e^2 -3\right) \tan ^{-1}\left(\frac{e}{\sqrt{1-e^2}}\right)+  \frac{36 d}{a}e
     \sqrt{1-e^2}}+ 24 e^3 \cos (q)+  24 e^3\right)},
     \label{eqn:17}
  \end{equation}\\
  \\
  
  \text{Prolate Spheroids:}
 
  \begin{equation}
   i \omega _{\pm}(q,e) = \pm \frac{1}{8 \sqrt{3} \pi e^{3/2} } {\sqrt{\sin ^2\left(\frac{q}{2}\right) \left({\frac{6 d}{a} \left(e^2-3\right) \ln \left(\frac{1+e}{1-e}\right)+ \frac{36 d}{a} e}  + 24 e^3 \cos (q)+24 e^3 \right)}}.
   \label{eqn:18}
   \end{equation}
   \\
  \end{widetext}
  
  \begin{figure}[b]
          \begin{center}
          \includegraphics[width=8.5cm]{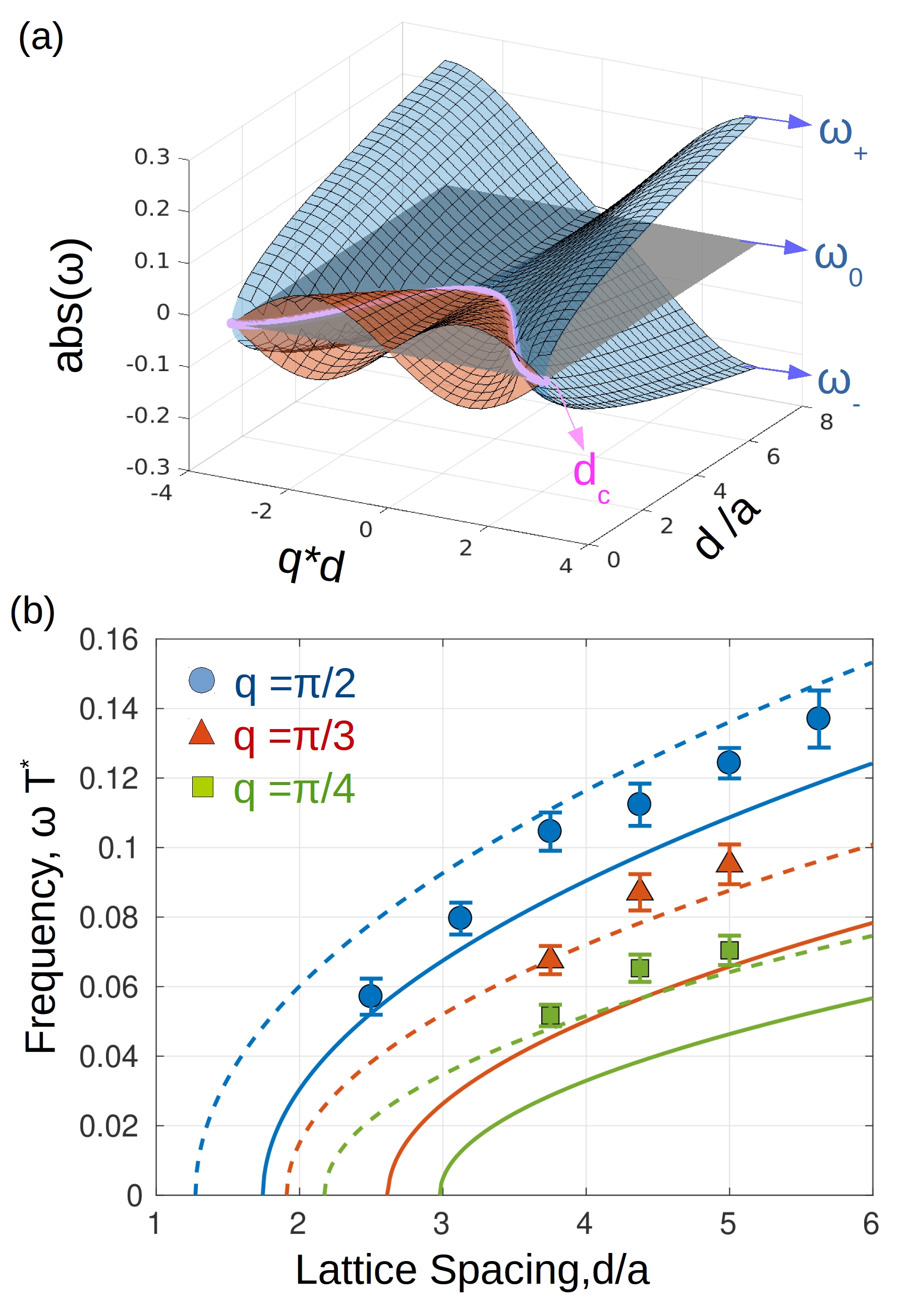}
          %\hspace{0cm}\newline
          \caption{\label{Fig5A}\textbf{Spectrum}:(a) The branches of the spectrum. The absolute value of the complex frequency is shown as a function of the nondimensional wavenumber $q$ and spacing $d$. The blue surface is for purely real frequencies (the neutrally stable modes) and the red surface is for purely imaginary frequencies (clumping instability). The boundary between the stable and unstable regimes is shown by the magenta curve $d_{c}$. (b) Measured non-dimensional frequency $\omega T^{*}$ (symbols) for various wavenumbers $q$ plotted against the lattice spacing $d/a$. The curves are predicted by our nearest-neighbour theory for oblate spheroids, solid lines for the experimental eccentricity $e=0.9922$ and dashed lines for the limit of zero thickness $e=1$.}
          \end{center}
    \end{figure}
 
 In the limit of  $e \to 0$, the dispersion relations for both oblate \eqref{eqn:17} and prolate \eqref{eqn:18} spheroids converges to $i \omega_{\pm} = \pm {|\sin(q)|}/4\pi$, which is just the Crowley instability for spheres \cite{crowley}. For $e \neq 0$, defining the nondimensional quantity $\tilde{d} \equiv 2 d \alpha(e) / 3 a$, gives a universal condition for stability:
 \begin{equation}
 \tilde{d} \geq \cos^{2} \frac{q}{2},
 \label{eqn:19}
 \end{equation}
 so that $\tilde{d} = q \cos^{2}/2$ defines the stability boundary in the $\tilde{d}$-$q$ plane, separating the regime of kinematic waves (blue) from the clumping instability (red) as shown in the phase diagram of Figure \ref{Fig4}(b). In general, for a uniaxial apolar shape, $\alpha$ is a constant parameter \cite{HB1} which can be determined by experimentally measuring the lateral drift of an isolated settling object. From \eqref{eqn:16} it follows that the above instability boundary and consequent dynamics is universal across all axisymmetric apolar shapes, when $d$ is rescaled by $2\alpha/3a$.
 
 This prediction agrees well with our experimental data shown by the red and blue circles, where we have initialised the lattice at those points in the $\tilde{d}$-$q$ plane.  The outcome of any given experiment is identified as being wave-like or clumping by considering the early stages of the time-dependence of $\braket{\rho(t) \delta\rho(t=0)}$, normalized amplitude of the density autocorrelation, which is measured by projecting the particle number density $\rho(t) = \sum_{m=1}^{N} \delta (x - x_{m}(t))/N$, on the initial density fluctuation $\delta\rho(t=0)$ of the lattice.  We obtained $\delta\rho(t=0)$ by fitting a sine to the initial horizontal displacement perturbation $u_{x}(t=0)$, and shifting in phase by $\pi/2$.  
 This is shown in the inset to Figure \ref{Fig4}(b), where some curves increase in amplitude, and others decay. At later times, even in the wave-like regime, the perturbation becomes very non-sinusoidal, as nonlinear effects become prominent.  
 
 More specifically, the limit of disks with zero thickness ($e \to 1$ for oblate spheroids), produces the mode structure shown in Figure \ref{Fig5A}(a):
 \begin{equation}
 i \omega_{\pm}(q) = \pm \frac{1}{4\pi} \sqrt{\sin ^2\left(\frac{q}{2}\right)\left(-\frac{d\pi}{2a} + 4\cos^{2} \frac{q}{2} \right)},
 \label{eqn:20}
 \end{equation}
which gives neutrally stable modes when the lattice spacing $d> 8 a \cos^{2} (q/2) / \pi$ and clumping instability otherwise. This prediction is compared with experimental data for the frequency in Figure \ref{Fig5A}(b) for various $q$ and $d$. We show solutions corresponding both to zero thickness, as well as for the ellipsoid with $2a$ and $2b$ corresponding to the diameter and thickness of our disks.
 
In the long wavelength limit $q \to 0 $, \eqref{eqn:20} reduces to the dispersion relation \eqref{eqn:8} predicted by that we showed in the previous section based on symmetry arguments. The mobility coefficients for disks are determined to be: $\lambda_{1}\lambda_{2} = -1/16\pi^{2}$ and $\alpha \gamma = d/128a\pi$.
 
The limit for needles of zero thickness (prolate spheroids with $e \to 1$) is not well defined, but the dispersion relation for rods of small thickness $2b$ and length $2a$, to leading order in $1-e$ is:
 \begin{equation}
  i \omega_{\pm}(q) = \pm \frac{1}{4\pi} \sqrt{\sin ^2\left(\frac{q}{2}\right)\left(\frac{3d}{a} -\frac {2d}{a} \ln\left(\frac{2a}{b}\right) + 2\cos^{2} \frac{q}{2} \right)}.
  \end{equation}
Note that the gapless feature ($\omega \to 0$ as $q \to 0$) of the modes \eqref{eqn:17} - \eqref{eqn:18} is tied to the conservation of total ``momentum'' $\sum_n \theta^n$ and the breaking of continuous translational symmetry by the lattice. This makes the lattice of orientable apolar objects resemble masses-and-spring chain as seen above \eqref{eqn:13}-\eqref{eqn:15}, with corresponding soft modes in the form of waves in displacement and orientation. Although the lattice was not formed by a phase transition to an ordered state, our imposition of an array structure on a translation-invariant background means that only the relative positions of disks matter, so the displacement field behaves like a true ``broken-symmetry'' mode \cite{martin}.

 \subsection{ Waves and non-modal growth}
 We show below that our system of sedimenting array of disks exhibits a special feature, since its dynamical matrix is non-normal. The short time behaviour of a system with a nonnormal dynamical matrix can be completely different from what one would expect from the exponential evolution of the eigenmodes \cite{schmid,bale}. In particular, even when all the eigenvalues show negative or zero growth rate, disturbances can grow algebraically for some time \cite{trefethen}. The quantum of growth depends on the operator itself, and on the configuration of the initial perturbation. For small growth, the system will relax at large times to the behaviour expected from the least stable eigenmode. When the transient growth is significant, however, the system is ultimately pushed into the non-linear regime. In hydrodynamic stability problems, especially in shear flows \cite{couette}, transition to turbulence through the algebraic growth route is quite common. However, experimental quantification of algebraic growth is extremely difficult. The present work offers a rare quantitative comparison of transient growth in theory and experiment.
 
   \begin{figure}[t]
            \begin{center}
            \includegraphics[width=8.5cm]{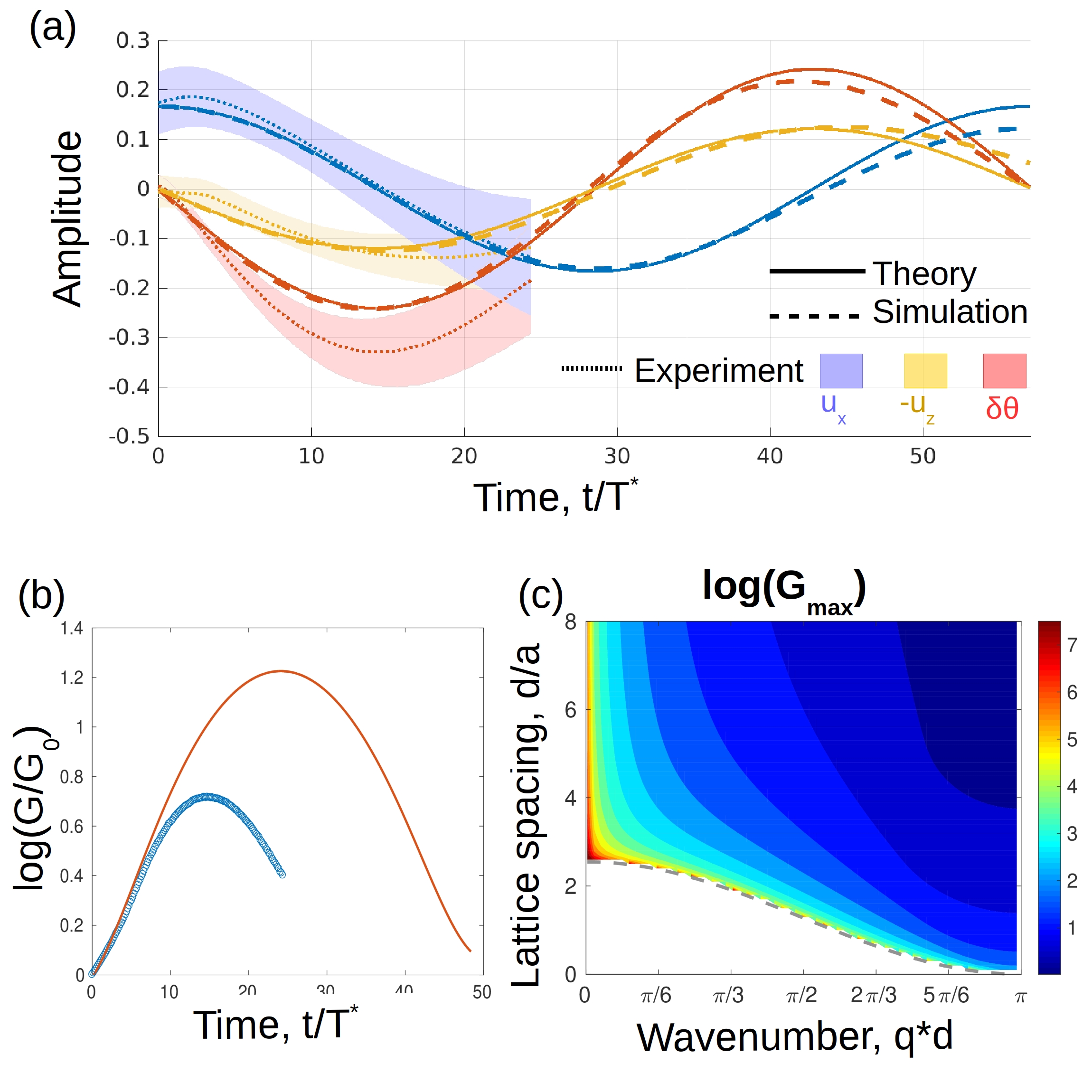}
            %\hspace{0cm}\newline
            \caption{\label{Fig5B}\textbf{Non-modal growth}:(a) The experimental amplitude for the angle (red patch), horizontal perturbation (blue patch) and vertical perturbation (yellow patch) compared against theory (solid lines) and far-field simulation of oblate spheroids in limit of zero thickness (dashed lines), for the stable case of $q=\pi/2$ and $d/a = 3.75$. The extent of the shaded region shows the corresponding error in measurement of amplitude.(b) The nonmodal growth plotted for the stable case $q=\pi/2$ and $d/a = 3.75$ in experiment (blue), compared with the simulation for the initial condition that gives maximum gain (red) in $t=14 T^{*}$, where $G_{0}$ is the initial amplitude.(c) The log of ratio $G/G_0$,  of maximum amplitude $G$ of the nonmodal perturbation to the initial amplitude $G_0$ depicted for the neutrally stable regime in the $q*d-d/a$ plane. }
            \end{center}
      \end{figure}
 
 For disks, the eigenfunctions of the dynamical matrix \eqref{eqn:16} can be used to construct the solution for experimental initial conditions. The  eigenvectors corresponding to the eigenvalues $(0, - i \omega, i \omega)$, where $\omega \equiv \omega_{+}$ \eqref{eqn:20}, are respectively given by the columns of the matrix
 \begin{equation}
 \left(
 \begin{array}{ccc}
  0 & {i\omega \pi \csc^{2}\left(\frac{q}{2}\right) } & {-i\omega \pi \csc^{2} \left(\frac{q}{2}\right)}\\
  -i \frac{d \pi}{8a}    \csc (q) & -\frac{1}{2} i \cot \left(\frac{q}{2}\right) &
    -\frac{1}{2} i \cot \left(\frac{q}{2}\right) \\
  1 & 1 & 1 \\
 \end{array}
 \right).
 \end{equation}
 The experimental initial perturbation $(u_{x}, u_{z}, \delta\theta) = (\epsilon, 0,0)$ in the neutrally stable regime gives waves of displacement and orientations [see Appendix C]:
 \begin{equation}
 u^{n}_{x}(t) = \epsilon \sin( q n) \cos( \omega t),
 \end{equation}
 \begin{equation}
 u^{n}_{z}(t) = \frac{\epsilon \sin(q)}{ 4\omega\pi} \cos(q n) \sin (\omega t), 
 \end{equation}
 \begin{equation}
  \delta \theta^{n}(t) =  -\frac{\epsilon\sin^{2}(q/2)}{\omega\pi} \sin(q n) \sin (\omega t).
 \end{equation}
 The time-dependence predicted here is in good agreement with the experimental data shown in Figure \ref{Fig5B} (a) for $q=\pi/2$ and $d/a=3.75$. This result is also compared to the numerical integration of the far field equations with periodic boundary conditions [as discussed in Appendix D], remaining in the limit of small $a/d$ but retaining nonlinearities to one further order in $u/r$ than in equations \eqref{eqn:9} - \eqref{eqn:12} [see Supplementary video 6]. 
  
  The  non-modal nature of our dynamical matrix $A$ \eqref{eqn:16} ($AA^\dagger \ne A^\dagger A$)  gives rise to non-orthogonal eigenvectors, resulting in non-modal growth of perturbations. So even in the `stable' regime of the phase diagram, perturbations show transient algebraic growth. If the transient amplitude is large enough, nonlinear growth takes over, as in our experiment. In our far-field numerical solution on the other hand, we have the facility to reduce the initial amplitude so much that despite transient growth the system remains linear. We observe significant nonmodal growth for the neutrally stable mode of experimental perturbations [see Fig \ref{Fig5B} (b)]. To quantify the nonmodal growth in $q-d$ plane, we calculate the norm of $\exp[\vect{A}s]$ for all times $s$, and calculate the maximum amplitude $G_{max}$ attained by the perturbation over all $s$, which is finite in the stable regime and depends on wavenumber $q$ and lattice spacing $d$ [see Fig \ref{Fig5B} (c)]. The singular value decomposition of $\exp(\vect{A} s)$ provides the initial condition which gives maximum nonmodal growth, which we compare against the experimental perturbation using far-field simulation of oblate spheroids in the limit of disk $e \to 1$ [see Fig \ref{Fig5B} (b)]. Further, our numerical study of the far-field equations [see Appendix D] shows that the observed disruption of the lattice in the stable regime results from amplification of the experimental noise in the initial orientations [see Supplementary videos 3 \& 7]. 
  
  We see thus that even in the regime where the orientational degree of freedom defeats the Crowley mechanism, and linear stability predicts waves, transient growth ultimately triumphs. An array of sedimenting spheroids is thus disrupted at all $q$ and $d$. In our numerical study with periodic boundary conditions we are able to observe the waves and delay the onset of nonlinearity by reducing the amplitude of the initial perturbations, unlike in the experiments where there are inevitable imprecisions in the initial conditions.
    
\section{Conclusions} 
The many-body physics of collective sedimentation holds many challenges and provokes many debates \cite{SR1,guazzelli}, which we must consider anew if we are to understand the role of internal degrees of freedom arising from particle shape. We study the role of particle orientation in the minimal setting of a one-dimensional lattice of Stokesian settling disks, and show the existence of two regimes of dynamical behaviour, as a function of lattice spacing and perturbation wavenumber. One of these is an extension of Crowley's clumping instability \cite{crowley} to non-spherical particles. The second is a hitherto unknown state of orientation and displacement waves, where the drift and mutual interaction of the disks overcome the clumping instability. We thus identify an unexpected mechanism to resist instabilities that were identified for spheres fifty years ago \cite{crowley}, were elaborated into parallel ideas about spheroids in a landmark paper thirty years ago \cite{KS}, and opened new directions in nonequilibrium statistical mechanics twenty years ago \cite{LR}. This competition between orientation and clumping in spheroids is related to an effect predicted for polar particles \cite{witten2}, and suggests a new consideration that must be included in the statistical theory of Koch and Shaqfeh \cite{KS}.  Further, we show that the momentum-like character of the particle orientation, seen earlier in pair of settling disks \cite{chajwa}, plays a crucial role in the collective dynamics of the disk array. The conservation of total ``momentum'', in conjunction with broken continuous translational invariance, yields gapless modes ($\omega \to 0$ as $q \to 0$) in this driven dissipative system.

The wave-like regime is unusual in that we predict, and observe experimentally, large transient growth that ultimately destabilizes the lattice, through nonlinear effects arising from the amplification of initial experimental error in release. Thus, the lattice is nonlinearly unstable over the entire $q-d$ plane, but due to two very different mechanisms. This unusual mechanism for nonlinear instability, namely, transient algebraic growth of perturbations in a linearly stable regime, should be of relevance in many other dissipative dynamical systems, but is not as widely known as it should be as not many examples have been identified and studied in the laboratory. We hope our experimental findings and theory on this remarkable effect will stimulate others to seek this mechanism in systems where the cause of long-term instability is ascribed to unidentified drifts or noise sources. The fact that our calculation, and the accompanying numerics, allow us to capture both the mode-structure and the growth of perturbation amplitude reassure us that our numerical model can in future be used to gain a comprehensive understanding of the unstable regime, and of other lattice configurations. 

There is no evolution of the angles when disks are globally rotated, thanks to the orientation-independence of the gravitational energy of \textit{apolar} shapes. Objects with \textit{polar} shape will have a preferred orientation in a gravitational field \cite{witten1, Ekiel, conway}, hence a damping of the ``momentum'' corresponding to $\sum_n\theta^n$, at zero wavenumber, and therefore an overdamping of the wavelike modes at small wavenumber. 

The dynamics of sedimenting lattices of more particles with complex shapes having non-zero polarity and chirality \cite{krapf} remains open to investigation and is expected to show behaviours distinct from orientable shapes discussed here. Finally, sedimenting objects in natural world, like flakes of clay particles in river beds or red blood cells settling in plasma are disk-like \cite{clay,ESR}, and we believe our work offers a useful building block to  understand the role of particle shape in these complex fluid-mechanical phenomena.

\medskip

\begin{acknowledgments}
RC and RG acknowledge support of the Department of Atomic Energy, Government of India, under project number 12-R\&D-TFR-5.10-1100. SR was supported by a J C Bose Fellowship of the SERB (India) and by the Tata Education and Development Trust, and acknowledges an Adjunct Professorship with TIFR. NM was supported through NSF DMR 1905698.
\end{acknowledgments}

\begin{appendices}
\appendix
\section{EXPERIMENTAL METHODS}
\subsection{Initial perturbations}
\begin{figure}[b]
      \begin{center}
      \includegraphics[width=8.5cm]{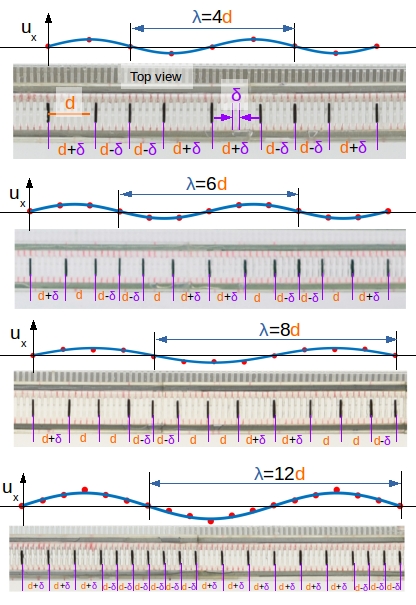}
      %\hspace{0cm}\newline
      \caption{\label{fig:AA1}\textbf{Release mechanism}: Top view of the release mechanism shown before release, with the disks loaded in the slots. The centre-to-centre distance between adjacent slots is $\delta$ and the lattice spacing of reference lattice is $d$. Initial horizontal positional perturbation $u_{x}$ are shown by fitting a sine function (blue curve) to the measured perturbations (red dots) for the wavelengths $\lambda = 4d, 6d, 8d$ and $12d$. }
      \end{center}
\end{figure}
\begin{figure}[h]
      \begin{center}
      \includegraphics[width=8.5cm]{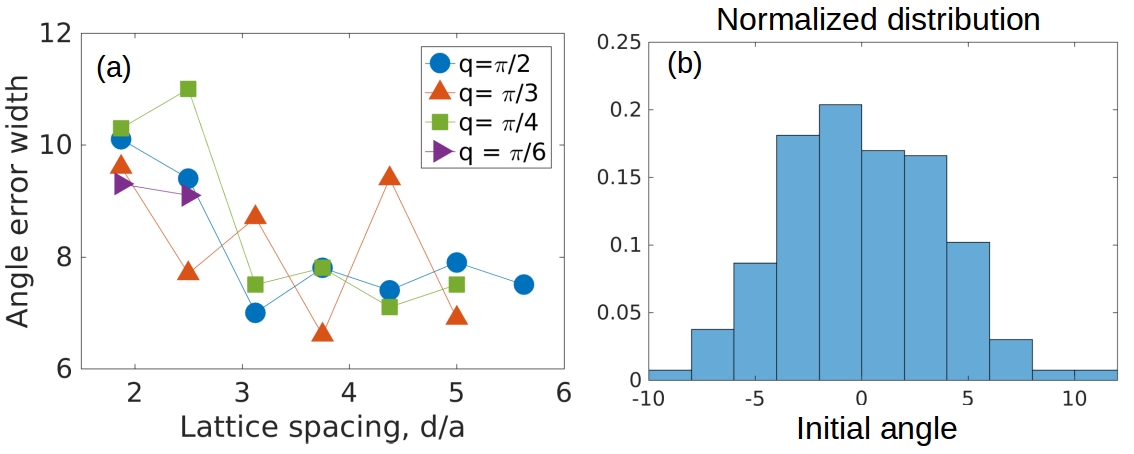}
      %\hspace{0cm}\newline
      \caption{\label{fig:AA3}\textbf{Error in initial orientation}:  (a) The error is quantified by $0.5 \times $(maximum - minimum) of the initial angles in any given run. This measure of the width in release angle is plotted here as a function of lattice spacing for  $qd = \pi/2, \pi/3, \pi/4$ and $\pi/6$. (b) a combined distribution of, initial angles of several experimental runs for a fixed $qd= \pi/2$ and $d=3.75 a$.}
      \end{center}
\end{figure}
The control parameters in experiments are the lattice spacing of the reference lattice and wavelength of the initial horizontal positional perturbation around this reference state. This was achieved using 3D printed stacks of rectangular slots of width $0.125$ cm and height $0.90$ cm, with disks of thickness $0.1$ cm and diameter $0.8$ are arranged in a periodic pattern initially as shown in figure. This perturbation was made as close to sinusoidal as possible within the constrain of discretization imposed by the slots [see Fig(\ref{fig:AA1})]. After arranging the disks across the total length of the release mechanism of $80$ cm, the disks were poked out gently while the whole mechanism was submerged roughly 3 cm below the surface of the fluid, to avoid any bubbles. After the disks were released out of the slots, they were measured to have a random orientation error sitting on the spatial perturbation which we imposed [see Fig(\ref{fig:AA3})]. This angular error corresponds to an error in horizontal spatial perturbation in $u_{x}$ of $\pm 0.03$ cm. This error in release plays a crucial role in disrupting the lattice at late times in the linearly stable, but transiently growing, regime.
\subsection{Measuring frequency}
The wave nature is evident in the dynamics of the orientations and positional perturbations which exhibit a quarter cycle of the wave with reasonable accuracy, before the non-linear instabilities kick in via an algebraic growth of perturbations, in contrast with the exponential growth of perturbations in the unstable regime. The positions and orientations of the disks are measured at each frame every 3 seconds by fitting an ellipse around each disk. The reference lattice is constructed by measuring the largest node to node distance in the initial condition. It is assumed that this reference lattice settles down vertically with the mean settling speed of the lattice and the orientation and positional perturbations are measured for the particles corresponding to each lattice point. 
\begin{figure}[h]
      \begin{center}
      \includegraphics[width=8.5cm]{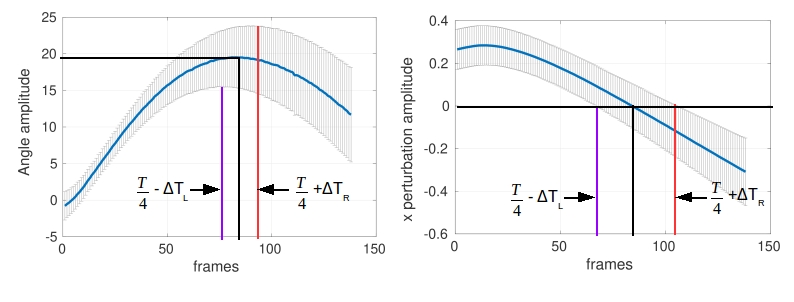}
      %\hspace{0cm}\newline
      \caption{\label{fig:AA2}\textbf{Frequency measurement}: Amplitude of the fitted sine wave plotted for the angle and the horizontal positional perturbations, along with the residual error of the fit shown as vertical error-bar at each frame. At quarter time period of the wave, the amplitude of angle peaks and the horizontal perturbation crosses $y=0$ axis, consistent with our theory. }
      \end{center}
\end{figure}
A sine wave is fitted in the perturbation with specified wavenumber at each frame and the amplitude of the fitted wave is measured at each time step [see Fig(\ref{fig:AA2})]. The residual of this fit gives error in frequency measurements as the amplitude for $u_{x}$ and $\theta$ is plotted as a function of time. 

\section{CONSTRUCTING MOBILITY USING SYMMETRIES}
The gradient expansion of the translational mobility $\bsf{M} (\nabla \vect{u}, \vect{K}, \nabla \vect{K})$ and rotational mobility $\bsf{N}  (\nabla \vect{u}, \vect{K}, \nabla \vect{K})$, to leading orders in gradients gives 
\begin{align}
\bsf{M} = &  \, \bsf{M}_{0} \, + \,  \bsf{M}_{1} \nabla \vect{ u} \, + \, m_{2} \,  \vect{K} \vect{K}  + \, \mathcal{O}(\nabla \nabla \vect{u}) \,+   \, \mathcal{O}(\nabla \nabla \vect{K})  
 \label{eqn:3}
\end{align}
\begin{align}
\bsf{P} \cdot \bsf{N} & =  n_{1}\, {\boldmath{\epsilon}}  \cdot \vect{K} \cdot ( {\boldmath{\epsilon}} \cdot \nabla \nabla \cdot \vect{u} )   \, + \, n_{2}\, \, \bsf{P} \cdot \vect{K}\, \nabla \nabla \cdot \vect{ u} \quad \quad \nonumber \\
& + \, n_{3} \, \, \bsf{P} \cdot \nabla \vect{K} \, + \mathcal{O}(\nabla \vect{u} \nabla \vect{u}) \, + \, \mathcal{O}(\nabla \nabla \vect{K})
\label{eqn:4}
\end{align}
Here $\bsf{M}_{0}$ is the mobility of the undistorted lattice and such a term in not allowed in $\bsf{N}$ due to symmetry under $\vect{K} \to -\vect{K}$; and $\bsf{P} \equiv \bsf{I} - \vect{K}\vect{K}$ is the projector transverse to the unit vector $\vect{K}$. In the first term of \eqref{eqn:4}, {\boldmath{$\epsilon$}} is the Levi-Civita tensor. Retaining only those terms that are allowed by the symmetries, leads to the "hydrodynamic" equations for the displacement field $\vect{u}$ and orientation field $\vect{K}$ in one dimension $x$ by dropping $z$ derivatives \eqref{eqn:5} - \eqref{eqn:7}.
\section{WAVE SOLUTIONS FOR SPHEROIDS}
The eigenfunctions corresponding to the eigenvalues $(\lambda_{1},\lambda_{2},\lambda_{3}) = (0, - i \omega, i \omega)$, where $\omega \equiv \omega_{+}$ is form \eqref{eqn:17}- \eqref{eqn:18}, are given by $\vect{v}_{1}$, $\vect{v}_{2}$ and $\vect{v}_{3}$ respectively 
\begin{equation}
\vect{v}_{1} = \left(
\begin{array}{c}
0\\
-i \frac{2 d \pi \alpha(e)}{3a}    \csc (q)\\
1\\
\end{array}
\right),
\vect{v}_{2} = \left(
\begin{array}{c}
{i\omega \pi \csc^{2}\left(\frac{q}{2}\right) } \\
-\frac{1}{2} i \cot \left(\frac{q}{2}\right)\\
1\\
\end{array}
\right)
\nonumber
 \end{equation}
 \begin{equation}
 \vect{v}_{3} = \left(
 \begin{array}{c}
 {-i\omega \pi \csc^{2}\left(\frac{q}{2}\right) } \\
 -\frac{1}{2} i \cot \left(\frac{q}{2}\right)\\
 1\\
 \end{array}
 \right)
 \end{equation}
 giving the solution as a real part
 \begin{equation}
 \vect{X}(t) = \sum^{3}_{i =1} \frac{a_{i}}{2} (\vect{v_{i}} e^{iqn} e^{\lambda_{i}t} + \vect{v^{*}_{i}} e^{-iqn} e^{\lambda^{*}_{i}t} )
 \label{B2}
 \end{equation}
 here $\vect{X}^{n} = (u^{n}_{x}, u^{n}_{y}, \delta \theta^{n}_{q})^{\intercal} $ . For $\vect{a} \equiv (a_{1}, a_{2}, a_{3})$, \eqref{B2} becomes $\vect{X}^{n}(t) = \vect{B} \cdot \vect{a}$, where in the stable regime
 
 \begin{widetext}
 \begin{equation}
 \vect{B} =
  \left(
  \begin{array}{ccc}
   0 & {-\omega \pi \csc^{2}\left(\frac{q}{2}\right) \sin(q n -\omega t) } & {\omega \pi \csc^{2} \left(\frac{q}{2}\right)  \sin(q n + \omega t) }\\
   \frac{2 d \pi \alpha(e)}{3a}    \csc (q) \sin(q n) & \frac{1}{2} \cot \left(\frac{q}{2}\right) \sin(q n -\omega t)  &
     \frac{1}{2}  \cot \left(\frac{q}{2}\right)  \sin(q n + \omega t) \\
   \cos(q n) & \cos(q n -\omega t) & \cos(q n + \omega t) \\
  \end{array}
  \right)
 \end{equation}
 \end{widetext}
 
 The coefficients $\vect{a}$ can be determined from the initial condition, $\vect{a} = \vect{B}^{-1} \cdot \vect{X}^{n}|_{t=0} $. Our experimental initial condition is $\vect{X}^{n}(t=0) = (\epsilon \sin(q n), 0,0 )^{\intercal}$, making $\vect{a} = \frac{\epsilon \sin^{2}(q/2)}{ 2\omega\pi} ( 0, -1,1)^{\intercal} $, which gives the wave solution in stable regime
 \begin{equation}
 u^{n}_{x}(t) =  \epsilon \sin(q n) \cos(\omega t)
 \end{equation}
 \begin{equation}
 u^{n}_{z}(t) = \frac{\epsilon \sin(q)}{ 4\omega\pi} \cos(q n) \sin (\omega t) 
 \end{equation}
 \begin{equation}
 \delta \theta^{n}(t) =  -\frac{\epsilon\sin^{2}(q/2)}{\omega\pi} \sin(q n) \sin (\omega t) 
 \end{equation}
 Note that the dependence on eccentricity of the spheroids enters through $\omega$ from \eqref{eqn:17} \& \eqref{eqn:18}. In the unstable regime the eigenvalues $(0, -\lambda, +\lambda)$ are real, giving hyperbolic functions in the time dependence of the solution
 \begin{equation}
  u^{n}_{x}(t) =  \epsilon \sin(q n) \cosh(\lambda t)
  \end{equation}
  \begin{equation}
  u^{n}_{z}(t) = \frac{\epsilon \sin(q)}{ 4\lambda\pi} \cos(q n) \sinh (\lambda t) 
  \end{equation}
  \begin{equation}
  \delta \theta^{n}(t) =  -\frac{\epsilon\sin^{2}(q/2)}{\lambda\pi} \sin(q n) \sinh (\lambda t) 
  \end{equation}
 
 \section{FAR-FIELD SIMULATIONS WITH PERIODIC BOUNDARIES}
 To understand the non-linear dynamics of disks in $(x,z)$ plane, we numerically analyse the equations of motion for spheroids in the limiting case of disks $e \to 1$, to leading order in $\mathcal{O}(a/r)$, by pairwise addition of hydrodynamic interactions using the method of reflections \cite{kim2}. We simulate the following equations for positions $(x_{n},z_{n})$ and orientations $\theta_{n}$ of the $n^{th}$ spheroid with using fourth order Runge-Kutta method:
\begin{equation}
\frac{d x_{n}}{dt} =  \sin 2\theta_{n} \frac{d }{64 a} - \sum_{m \neq n}^{N} \frac{(x_{n} - x_{m})(z_{n} - z_{m})}{8 \pi \, {r^{3}}_{mn}}
\end{equation} 
 \begin{equation}
 \frac{d z_{n}}{dt} =  (\sin ^2\theta_{n} -3 ) \frac{d }{32 a}  - \sum_{m \neq n}^{N} \left[\frac{1}{8 \pi \, r_{mn}}
 + \frac{(z_{n} - z_{m})^{2}}{8 \pi \, {r^{3}}_{mn}} \right]
 \end{equation} 
  \begin{align}
  \frac{d\theta_{n}}{dt}& = \sum_{m \neq n}^{N}  \frac{(x_{n} - x_{m})}{8 \pi \, {r^{3}}_{mn}}  \nonumber \\
  &  - \sum_{m \neq n}^{N} \frac{3 (z_{n} - z_{m})}{8 \pi \, {r^{5}}_{mn}}  \left[(x_{n} - x_{m})\cos \theta_{n} + (z_{n} - z_{m})\sin \theta_{n} \right] \, \times \nonumber \\
    & \left[(z_{n} - z_{m})\cos \theta_{n} - (x_{n} - x_{m})\sin \theta_{n}\right]
  \end{align}
 In the nearest-neighbour approximation, number of interacting neighbours $N$ truncates the spatial summation over $m$, making implementation of periodic boundaries straightforward.  Note that the above equations are non-dimensionalized using length scale $d$ and time scale $\mu d^{2} /F$. Also, the initial conditions are such that the orientation vector of all the spheroids lie in the $(x,z)$ plane and hence the resulting trajectories are confined to the same plane $y = 0$.
 
 \section{DETAILS OF SUPPLEMENTARY VIDEOS}
 \subsection{Video 1: Crowley's Mechanism} 
 Five spheres of diameter 0.6 cm prepared in an array perturbed around an equally spaced configuration with an interparticle spacing 1.5 $\pm$ 0.1 cm. The initial perturbation is like that of Fig\ref{Fig1} (b) with amplitude 0.25 $\pm$ 0.05 cm. Trajectories of the nodes of this perturbation is shown in red. The three-sphere dynamics at later times is expected to be chaotic \cite{janosi}.

 \subsection{Video 2: Linearly stable wavelike mode}
  Initial sinusoidal perturbation with $d=3.75a, qd = \pi/2$,  and amplitude $0.625 a$; and with trajectory of nodes shown by the dashed red lines. We zoom in on a region where initial errors in release were small, which shows a  half cycle of the wavelike oscillation in orientations and positions. More details are in Fig.\ref{Fig2}(b).
  
 \subsection{Video 3: Disruption of waves at late times}
 This video shows the late-time dynamics for the same $q$ and $d$ as in Video 2. Transient algebraic growth of the perturbations leads to nonlinear effects that disrupt the array. 
 
 \subsection{Video 4: Linearly unstable mode}
 Initial sinusoidal perturbation with $d=2.5a, qd = \pi/4$ and amplitude $0.625 a$. More details given in Fig.\ref{Fig3}(b).\\
 
 \subsection{Video 5: Clumping dynamics at late times.}
 Late time clumping behaviour of the perturbation with $d=1.875a, qd = \pi/6$ and amplitude $0.625 a$.  Trajectories of nodes are shown in red color.\\

 \subsection{Video 6: Numerical study of wave-like regime}
 Numerical integration of the non-dimensionalised far-field equations [see Appendix D] with initial sinusoidal perturbation of $q = \pi/2$. The interaction is cut-off beyond $1.5d$, such that only nearest neighbours interact hydrodynamically. The  region shown here is the same size as the experimental container, scaled by lattice spacing. The initial conditions is the same as in the experiment of Video 2, albeit with periodic boundary condition and no experimental error in initial condition [see Fig \ref{Fig5B} (a)].   
 \subsection{Video 7: Numerical study of wave-like regime with noisy initial conditions}
The initial condition is the same as in Video 6, but we add a random error in the initial orientations uniformly randomly distributed between $\pm 8 \deg$,  to reflect the measured experimental initial conditions of Video 3 [see Fig \ref{fig:AA3}].
\end{appendices}
 
%\pagebreak

\end{document}